\documentclass[%
 reprint,
superscriptaddress,
nobibnotes,
footinbib,
amsmath,amssymb,
prl,
nolongbibliography
]{revtex4-2}

\usepackage[Symbol]{upgreek}
\usepackage{graphicx}% Include figure files
\usepackage{dcolumn}% Align table columns on decimal point
\usepackage{bm}% bold math
\usepackage{hyperref}% add hypertext capabilities
\hypersetup{
     colorlinks   = true,
     linkcolor    = blue,
     citecolor    = blue,
     breaklinks   = true, % for arxiv to break long url in bibliography
}

\begin{document}

\preprint{APS/123-QED}

\title{First-Principles Electron-Phonon Interactions and Polarons\\ in the Parent Cuprate La$_2$CuO$_4$}

\author{Benjamin K. Chang}
\affiliation{Department of Applied Physics and Materials Science, and Department of Physics, California Institute of Technology, Pasadena, California 91125, USA}
\author{Iurii Timrov}
%\affiliation{Theory and Simulation of Materials (THEOS), and National Centre for Computational Design and Discovery of Novel Materials (MARVEL), \'{E}cole Polytechnique F\'{e}d\'{e}rale de Lausanne (EPFL), CH-1015 Lausanne, Switzerland}
\affiliation{Laboratory for Materials Simulations, Paul Scherrer Institut, 5232 Villigen PSI, Switzerland}
\author{Jinsoo Park}
\affiliation{Department of Applied Physics and Materials Science, and Department of Physics, California Institute of Technology, Pasadena, California 91125, USA}
\affiliation{Pritzker School of Molecular Engineering, University of Chicago, Chicago, Illinois 60637, USA}
\author{Jin-Jian Zhou}
\affiliation{School of Physics, Beijing Institute of Technology, Beijing 100081, China}
\author{Nicola Marzari}
\affiliation{Theory and Simulation of Materials (THEOS), and National Centre for Computational Design and Discovery of Novel Materials (MARVEL), \'{E}cole Polytechnique F\'{e}d\'{e}rale de Lausanne (EPFL), CH-1015 Lausanne, Switzerland}
\affiliation{Laboratory for Materials Simulations, Paul Scherrer Institut, 5232 Villigen PSI, Switzerland}
\author{Marco Bernardi}
\email{bmarco@caltech.edu}
\affiliation{Department of Applied Physics and Materials Science, and Department of Physics, California Institute of Technology, Pasadena, California 91125, USA}
%\affiliation{Department of Physics, California Institute of Technology, Pasadena, California 91125, USA}
%

\begin{abstract}
Understanding electronic interactions in high-temperature superconductors is an outstanding challenge. 
In the widely studied cuprate materials, experimental evidence points to strong electron-phonon ($e$-ph) coupling and broad photoemission spectra. Yet, the microscopic origin of this behavior is not fully understood. Here we study $e$-ph interactions and polarons in a prototypical parent (undoped) cuprate, La$_2$CuO$_4$ (LCO), by means of first-principles calculations. Leveraging parameter-free Hubbard-corrected density functional theory, we obtain a ground state with band gap and Cu magnetic moment in nearly exact agreement with experiments. 
This enables a quantitative characterization of $e$-ph interactions. Our calculations reveal two classes of longitudinal optical (LO) phonons with strong $e$-ph coupling to hole states. These modes consist of Cu-O plane bond-stretching and bond-bending as well as vibrations of apical O atoms.  
The hole spectral functions, obtained with a cumulant method that can capture strong $e$-ph coupling, exhibit broad quasiparticle peaks with a small spectral weight ($Z\approx0.25$) and pronounced LO-phonon sidebands characteristic of polaron effects. 
Our calculations predict features observed in photoemission spectra, including a 40-meV peak in the $e$-ph coupling distribution function not explained by \mbox{existing models.}
These results show that the universal strong $e$-ph coupling found experimentally in lanthanum cuprates is an intrinsic feature of the parent compound, and elucidates its microscopic origin.  
\end{abstract} 

\maketitle
%
%%% ------  INTRO  -----
%
Angle-resolved photoemission spectroscopy (ARPES) has provided ample evidence for broad spectral functions in several cuprate compounds~\cite{kim_anomalous_2002, shen_missing_2004, rosch_polaronic_2005, ronning_anomalous_2005, shen_angle-resolved_2007}. 
This spectral broadening has been associated with strong electron-phonon ($e$-ph) interactions and polaronic behavior in doped and undoped cuprates~\cite{shen_missing_2004, rosch_polaronic_2005, shen_angle-resolved_2007}. Existing models can account phenomenologically for the observed spectral broadening~\cite{shen_missing_2004, kim_anomalous_2002, sawatzky_testing_1989}. Yet, developing a deeper understanding based on rigorous theory and quantitative calculations has been difficult, mainly due to the strong electron correlations governing cuprate physics~\cite{gunnarsson_interplay_2008, keimer_quantum_2015}. 
\\
\indent
In parent (undoped) cuprate compounds, the strong Coulomb repulsion of localized Cu 3$d$ electrons induces an antiferromagnetic (AFM) Mott insulating ground state~\cite{lee_doping_2006, damascelli_angle-resolved_2003} which can be described qualitatively using Hubbard-like or $t$–$J$ models~\cite{zhang_effective_1988, PRX-tj}. These Hamiltonians can also be combined with model $e$-ph interactions to predict the broadening of electron spectral functions~\cite{rosch_electron-phonon_2004, mishchenko_electron-phonon_2004, rosch_polaronic_2005, rosch_dispersion_2005, makarov_polaronic_2015, shneyder_polaron_2020}. 
However, key microscopic quantities needed for a realistic description of $e$-ph coupling in most high-temperature superconductors remain unknown, including the strength of the $e$-ph interactions, their dependence on electron and phonon momenta, their effects on electron spectral functions, and which atomic vibrations dominate the coupling.
\\
\indent
Owing to recent progress, first-principles calculations are able to characterize $e$-ph interactions and electron spectral functions also in correlated metals and insulators~\cite{zhou_ab_2021, li_unmasking_2021, abramovitch_combining_2023}. 
For cuprates, such quantitative studies have so far focused on metallic (doped) compounds relying on the local-density approximation (LDA)~\cite{savrasov_linear_1996, heid_momentum_2008, giustino_small_2008}, which cannot correctly describe the Mott insulating ground state of parent cuprates. 
Recent work has studied parent cuprates using improved functionals~\cite{pokharel_sensitivity_2022, furness_accurate_2018,lane_antiferromagnetic_2018,ning_critical_2023} or Hubbard-corrected density functional theory (DFT$+U$)~\cite{wei_electronic_1994, anisimov_spin_1992, sterling_effect_2021} to obtain reliable ground state and phonon spectra~\cite{zhang_electron-phonon_2007, sterling_effect_2021}. 
Even within these improved schemes, $e$-ph interactions in parent cuprates remain unexplored.
\\
\indent
In this Letter, we show fully \textit{ab initio} calculations of $e$-ph interactions and electron spectral functions in a prototypical parent cuprate, La$_2$CuO$_4$ (LCO). We employ a combination of advanced first-principles techniques, including linear-response DFT$+U$ and recently developed treatments of anharmonic phonons~\cite{monacelli_stochastic_2021}, strong $e$-ph interactions, and polarons~\cite{zhou_ab_2016, zhou_predicting_2019}.
% with properties in excellent agreement with experiments, 
Starting from an accurate ground state, we show that $e$-ph interactions in LCO are governed by two families of longitudinal optical (LO) phonons with strong Fr\"ohlich-type coupling. 
These LO modes consist of stretching and bending of Cu-O bonds and vibration of apical O atoms. 
The computed valence band spectral functions exhibit significant peak broadening and renormalization, as well as pronounced phonon sidebands. These features are governed by the strongly-coupled LO phonons, with a smaller contribution from lower-energy polar modes. Our results provide a quantitative evidence for strong $e$-ph interactions and polarons in parent cuprates mediated by multiple optical phonons, thus deepening our microscopic understanding of cuprate physics beyond analytical models.  
\\
\indent
We compute the ground state of LCO in the low-temperature orthorhombic phase~\cite{reehuis_crystal_2006, tranquada_neutron_2007, yamaguchi_observation_1987} using collinear spin-polarized DFT$+U$ calculations in a plane-wave basis with the \textsc{Quantum ESPRESSO} package~\cite{giannozzi_quantum_2009,giannozzi_advanced_2017}. The Hubbard-$U$ parameter for the Cu $3d$ states is calculated (rather than fitted) self-consistently with the relaxed crystal structure~\cite{timrov_self-consistent_2021,timrov_hp_2022,Iurii_Hubbard}, thus removing any tunable parameter in our calculations.
We employ the \mbox{SSCHA} method~\cite{monacelli_stochastic_2021} to compute  effective harmonic phonon dispersions at 150~K, a temperature where LCO is in its antiferromagnetic (AFM) state~\cite{yamaguchi_observation_1987, keimer_neel1_1992}. 
The $e$-ph perturbation potentials, obtained on a coarse irreducible $\mathbf{q}$-point grid with DFPT$+U$~\cite{floris_hubbard-corrected_2020, giannozzi_advanced_2017}, include both the Kohn-Sham and Hubbard perturbation terms~\cite{zhou_ab_2021}. 
We use the \textsc{Perturbo} code~\cite{zhou_perturbo_2021} to compute the $e$-ph matrix elements~\cite{zhou_perturbo_2021, zhou_ab_2021}, $g^{\sigma}_{mn\nu}(\mathbf{k},\mathbf{q})$, and interpolate them using Wannier functions from \textsc{Wannier90}~\cite{mostofi_wannier90_2008}. 
(These matrix elements represent the probability amplitude for an electron in a Bloch state $\psi^{\sigma}_{n\mathbf{k}}$, with band index $n$, spin $\sigma$, and crystal momentum $\mathbf{k}$, to scatter into state $\psi^{\sigma}_{m\mathbf{k+q}}$ by emitting or absorbing a phonon with mode index $\nu$ and wave-vector $\mathbf{q}$~\cite{zhou_perturbo_2021, bernardi_first-principles_2016, giannozzi_advanced_2017}.)
The electron spectral functions are computed at finite temperature with a cumulant approach described in Ref.~\cite{zhou_predicting_2019}. This method can capture strong $e$-ph interactions and polaron effects, such as the broadening and weight renormalization of the quasiparticle peak and the emergence of phonon sidebands, as we have shown in recent studies on oxides~\cite{zhou_predicting_2019, zhou_ab_2021} and organic crystals~\cite{chang_intermediate_2022}. Additional computational details are provided in the Supplemental Material (SM)~\cite{supplemental_materials}.\\
\begin{figure}[t!]
\includegraphics[width=0.48\textwidth]{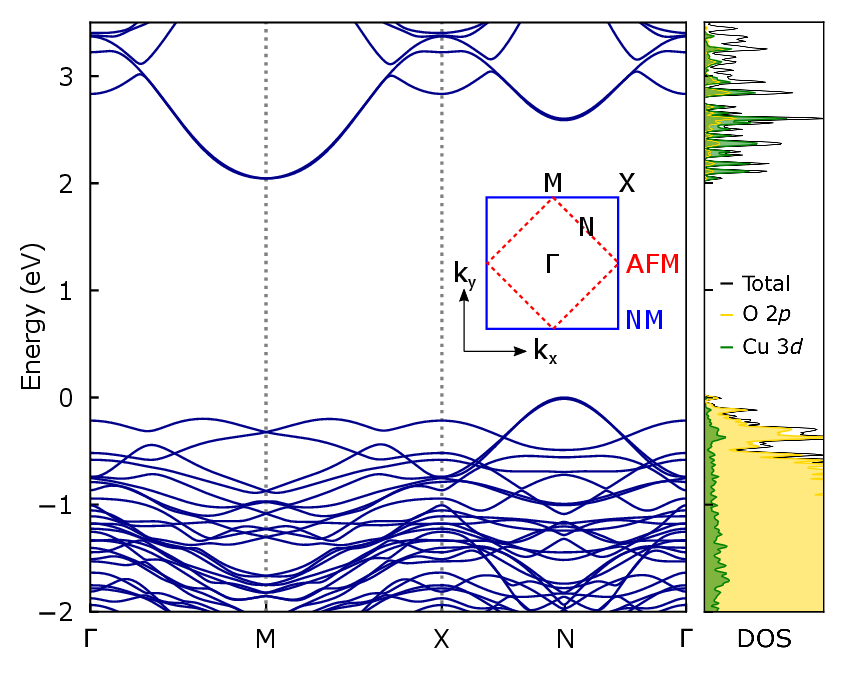}
\caption{Band structure of LCO in its AFM phase computed with DFT$+U$. The inset shows the in-plane Brillouin zone of the nonmagnetic (NM, blue) and  antiferromagnetic (AFM, red) phases, with high-symmetry points $\Gamma$, $M$, $N$ and $X$ labeled. The right panel gives the total density of states (DOS) and its contributions from O 2$p$ and Cu 3$d$ atomic states. The energy zero is set to the top of valence band.} 
\label{fig:band} 
\end{figure}
\begin{figure}[hbt!]
\includegraphics[width=0.48\textwidth]{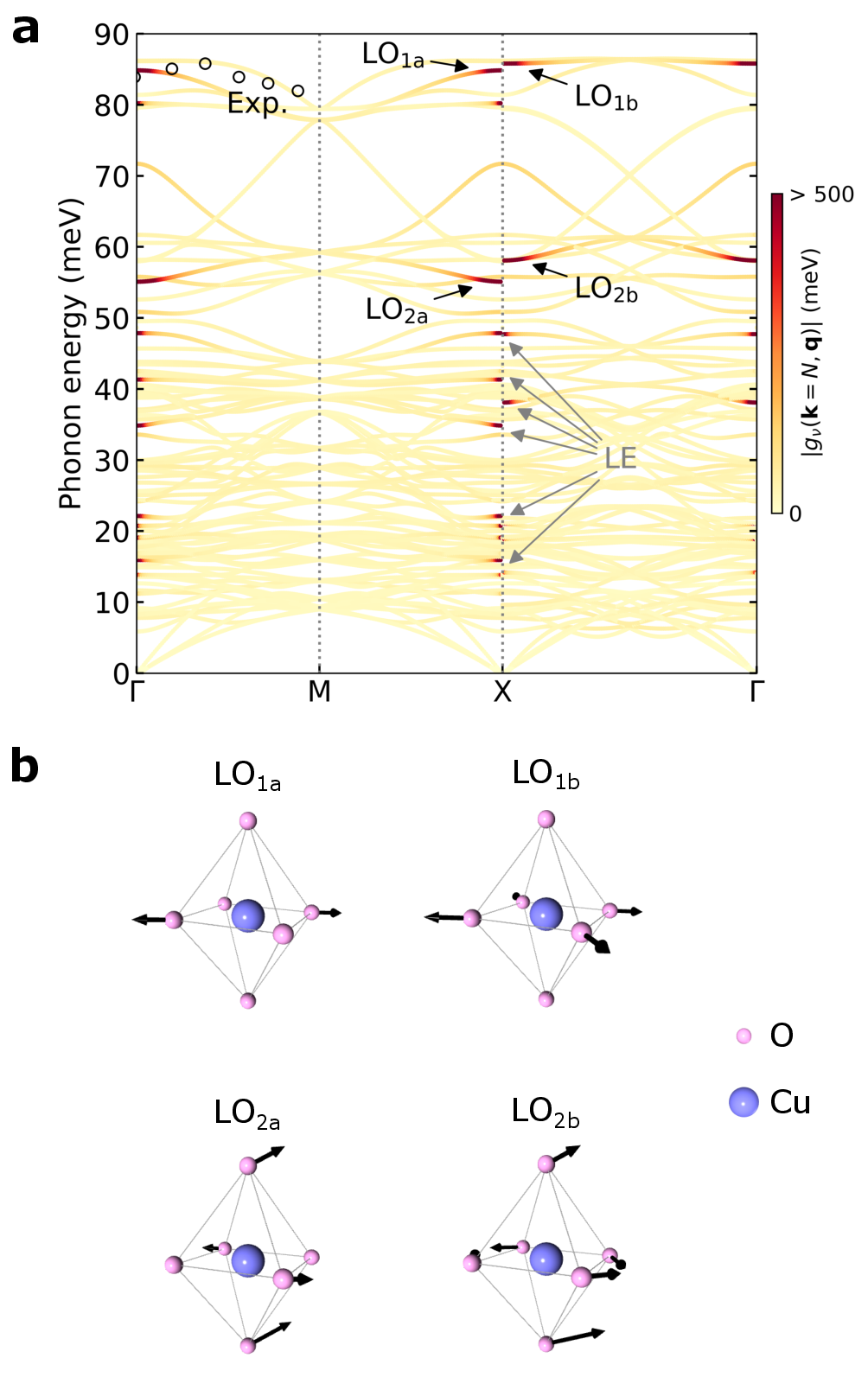}
\caption{(a) Calculated phonon dispersion overlayed with the $e$-ph coupling strength $|g_\nu(\mathbf{k},\mathbf{q})|$ computed at the nodal point $\mathbf{k}=N$. Experimental LO-mode energies are given for comparison with hollow circles~\cite{park_evidence_2014}. The four most strongly-coupled LO modes (LO$_{1a,b}$ and LO$_{2a,b}$) are indicated with black arrows, and lower-energy (LE) strongly-coupled modes with gray arrows. (b) Atomic displacements for the four strongly-coupled LO modes, with blue and pink spheres representing Cu and O atoms, respectively.}
\label{fig:phonon} 
\end{figure}
\indent
Figure~\ref{fig:band} shows the band structure of LCO computed with DFT$+U$. The valence band dispersion near the gap resembles conventional $t$–$J$ model results for insulating cuprates~\cite{wells_e_1995, sobota_angle-resolved_2021}, with the valence band maximum (VBM) at the nodal point $N$ [$\mathbf{k}\!=\!(\frac{\pi}{2}, \frac{\pi}{2})$] and additional lower-energy valleys near the antinodal point $M$ [$\mathbf{k}\!=\!(\pi, 0)$]. %(note that $X$ [$\mathbf{k}\!=\!(\pi, \pi)$] for the AFM phase is equivalent to $\Gamma$ for the nonmagnetic phase). 
The computed band gap is $E_g\!=\!2.04$~eV and the Cu magnetic moment is $\mu_{\mathrm{Cu}}\!=\!0.62\mu_{\mathrm{B}}$, in excellent agreement with the respective experimental values, $E_g^{\mathrm{exp}}\!=\!2.0$~eV~\cite{thio_determination_1990, ginder_photoexcitations_1988} and $\mu^{\mathrm{exp}}_{\mathrm{Cu}}\!=\!0.60$–$0.64\mu_{\mathrm{B}}$~\cite{kaplan_spin_1991,tranquada_neutron_2007} (see additional discussion in the SM~\cite{supplemental_materials}).
\\
\indent
Starting from this accurate ground state, we compute the phonon dispersions (inclusive of anharmonic effects) and map out the mode-resolved strength of the $e$-ph interactions in Fig.~\ref{fig:phonon}(a).
The experimental frequency of a Cu-O bond-stretching LO mode, measured by neutron scattering~\cite{park_evidence_2014} and given for comparison in Fig.~\ref{fig:phonon}(a), is in very good agreement with our calculations. 
As the high-symmetry $X$-point in the nonmagnetic (NM) phase is equivalent to the zone center ($\Gamma$ point) of the AFM phase, we find multiple discontinuities in the phonon dispersion near $X$ due to LO–TO splitting combined with orthorhomic $ab$-anisotropy, similar to the behavior observed in orthorhombic YBa$_2$Cu$_3$O$_{6+x}$~\cite{stercel_composition_2008, pintschovius_pronounced_2002}. 
We highlight the importance of temperature and anharmonic effects included in our SSCHA calculation; conversely, zero-temperature DFPT$+U$ phonons exhibit unphysical dynamical instabilities~\cite{supplemental_materials}. 
\\
\indent
Figure~\ref{fig:phonon}(a) also shows the $e$-ph coupling strength for each phonon mode~\cite{zhou_perturbo_2021}, $|g_\nu(\mathbf{k},\mathbf{q})|$, computed at the electron nodal point $\mathbf{k}\!=\!N$.
We identify two groups of LO phonons with strong Fr\"ohlich-type coupling~\cite{frohlich_electrons_1954}, named here LO$_1$ and LO$_2$, with respective energies  $\omega_{1}\!\approx\!85$~meV and $\omega_{2}\!\approx\!55$~meV.
We further distinguish between strongly coupled modes at opposite sides of the AFM zone center ($X$ point), labeling them with indices $a$ and $b$ respectively. 
At small wave-vector ($\Delta q\!\approx\!0.015$~\r{A}$^{-1}$) near $X$, 
these LO modes exhibit large $e$-ph coupling strengths, with values $|g_2|
\!\approx\!5.5$~eV for the LO$_{2a,b}$ and  $|g_1|\!\approx\!3.9$~eV for the LO$_{1a,b}$ phonons. 
These long-wavelength modes govern the $e$-ph physics in LCO as their $e$-ph coupling strengths are orders of magnitude greater than the Brillouin-zone average value (44~meV).
Such coupling strengths exceed those in strongly correlated metals (highest $|g|\!\approx\!100$~meV in Sr$_2$RuO$_4$)~\cite{abramovitch_combining_2023}, and have the same order of magnitude as the Fr\"ohlich coupling strengths in insulating oxides with polaron effects, including CoO and SrTiO$_3$~\cite{zhou_ab_2021, zhou_predicting_2019, zhou_electron_2018}.
\\
\indent
Notably, the energies of the LO$_1$ and LO$_2$ modes coincide with those of strongly-coupled phonons observed experimentally in several \textit{doped} cuprates, as evidenced by a universal kink in their quasiparticle (QP) band dispersion found in ARPES measurements~\cite{lanzara_evidence_2001, johnson_doping_2001, zhou_universal_2003}.
So far, these features have not been observed experimentally in undoped cuprates due to their broader ARPES spectra~\cite{ronning_anomalous_2005}.
\mbox{Our results} show quantitatively that strong $e$-ph interactions are already present at these energies in the parent cuprate LCO. 
\\
\indent
The four strongly-coupled LO modes are associated with O-atom vibrations, as shown in Fig.~\ref{fig:phonon}(b). The LO$_{1a}$ and LO$_{1b}$ modes correspond to vibrations in the CuO$_2$ plane consisting of oxygen breathing motions. 
These Cu-O bond-stretching modes have been shown to couple strongly with holes in neutron scattering measurements~\cite{mcqueeney_anomalous_1999} and are linked to Cu-O charge transfer~\cite{stercel_composition_2008} and formation of charge-ordered phases~\cite{pintschovius_pronounced_2002}. 
The LO$_{2a}$ and LO$_{2b}$ modes, on the other hand, involve both in-plane Cu-O bond-bending, with O atoms moving normal to the Cu-O bonds~\cite{ pintschovius_pronounced_2002}, and motion of the apical O atoms outside the CuO$_2$ planes~\cite{bianconi_determination_1996}. 
The $\sim$55~meV energy of the LO$_{\mathrm{2a,b}}$ modes is consistent with bond-bending modes measured in the doped LCO compound La$_{2-x}$Sr$_x$CuO$_4$ (LSCO)~\cite{mcqueeney_anomalous_1999}.
\\
\indent
We find additional lower-energy (LE) modes, with energies between 15$-$50 meV, whose $e$-ph coupling strengths are significant ($|g|\!\approx\!$ 1–2~eV at $\Delta q$ from $X$) but weaker than those of the LO$_{1,2}$ modes. These strongly-coupled LE modes consist mainly of Cu-O bond-bending and apical O vibrations, with modes at lower energies associated with higher-amplitude oscillations of Cu and La atoms. Animations for all strongly-coupled modes described above are given in the SM~\cite{supplemental_materials}.
Importantly, while Cu-O bond-stretching modes have received the most attention in the cuprates, our results show that bond-bending and apical oxygen motions possess the strongest $e$-ph coupling and govern $e$-ph physics in LCO.\\
\begin{figure}[t!]
\includegraphics[width=0.48\textwidth]{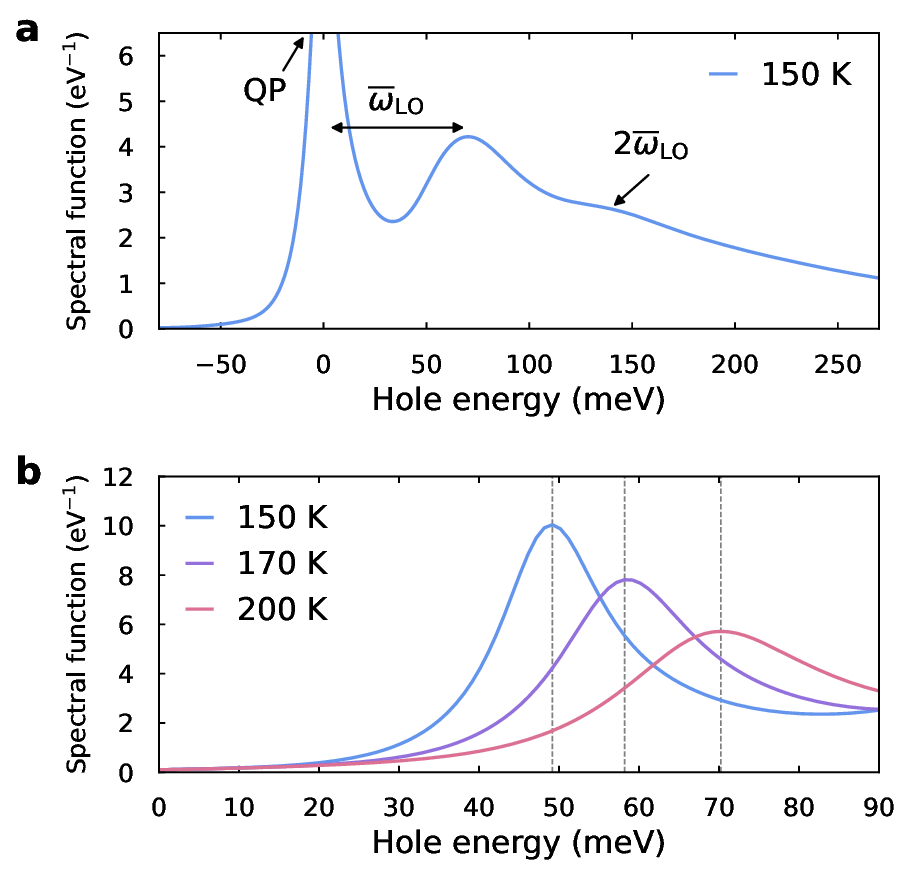}
\caption{(a) Spectral function of the VBM hole state in LCO computed at 150~K. The satellite peak is centered at $\overline{\omega}_{\rm LO}$, the average of the LO$_1$ and LO$_2$ phonon energies. A second overtone feature at $2\overline{\omega}_{\rm LO}$ is also shown. The energy zero is set to the QP peak energy. 
(b) Temperature-dependent spectral function of the VBM state computed at three temperatures, with the QP peak energies shown with dashed lines. The spectral functions are aligned at their energy onsets.}
\label{fig:spec} 
\end{figure}
%\\
\indent
The spectral functions exhibit clear signatures of strong $e$-ph coupling. Figure~\ref{fig:spec}(a) illustrates this result for the valence band maximum (VBM) hole state. 
The \mbox{$e$-ph interactions} broaden the QP peak and significantly decrease its spectral weight to $Z\!=\!0.25$.
In addition, we find a satellite peak at hole energy $\overline{\omega}_{\rm LO}\!\approx\!70$~meV accompanied by a less pronounced overtone at 2$\overline{\omega}_{\rm LO}\!\approx\!140$~meV relative to the QP peak~\footnote{Our results are given in terms of hole energies, which correspond to negative electron energies relative to the QP peak. Note also that phonon satellites appear at electron energies lower than the QP peak for $p$-doped materials, as we assume here; the sign of the satellite energy would be reversed in the $n$-doped case~\cite{zhou_predicting_2019}.}. 
These phonon sidebands form as a result of strong $e$-ph coupling with the LO$_1$ and LO$_2$ modes, whose average energy equals the satellite energy $\overline{\omega}_{\rm LO}\!\approx\!70$~meV. 
Other phonons with appreciable $e$-ph coupling, such as the LE modes, can also redistribute spectral weight and modify the spectral function. Yet, due to their lower energy and weaker $e$-ph coupling, their satellites carry less weight and merge into a broad incoherent background. 
\\
\indent
Our spectral function calculations can be viewed as a quantitative version of the Franck-Condon broadening (FCB) model, which describes the spectral functions in (un)doped cuprates as a superposition of multiple incoherent peaks~\cite{shen_missing_2004, kim_anomalous_2002, sawatzky_testing_1989}. 
The QP-peak broadening and renormalization in our calculations is consistent with both experimental results and the FCB model. Yet, our computed spectral functions exhibit well-defined QP peaks, which are typically missing in experiments on undoped cuprates~\cite{shen_missing_2004, kim_anomalous_2002}. 
We attribute this discrepancy to the small QP weight ($Z\!=\!0.25$), as predicted here, which makes the QP peak easily washed out in real samples by defect- and magnon-induced broadening not considered in this work~\cite{bohrdt_parton_2020, betto_multiple_2021, hamad_magnon_2021}. 
One difference with the FCB model is that the phonon sidebands decay rapidly with hole energy in our calculations, similar to previous results for large polarons in oxides~\cite{zhou_predicting_2019,zhou_ab_2021}, and are not visible beyond the second overtone at 2$\overline{\omega}_{\rm LO}$. In contrast, the FCB model predicts a series of intense satellites. 
\\
\indent
Our calculations can also explain an anomalous shift with temperature of the spectral function peak~\cite{kim_anomalous_2002}.
A model proposed by Kim \textit{et al.}~\cite{kim_anomalous_2002} predicts that the peak energy $E_p$ of the spectral function shifts with temperature according to $\Delta E_p \!\approx\! \pi k_{\mathrm{B}}\Delta T $, where $k_{\mathrm{B}}$ is the Boltzmann constant. This formula provides only a crude estimate $-$ for example, the measured temperature dependence is twice greater than predicted by this model in undoped Sr$_2$CuO$_2$Cl$_2$ and Ca$_2$CuO$_2$Cl$_2$ between 100$-$400~K. 
In contrast, our calculations on LCO can predict with a high accuracy the temperature dependence of the peak observed experimentally. %in doped cuprates.
Figure~\ref{fig:spec}(b) shows the computed spectral functions of the VBM hole state at three temperatures between 150$-$200~K. 
For an initial temperature of 150~K, increasing the temperature to 170~K and 200~K gives, respectively, a peak shift of 5 and 13~meV when using the simple model, versus 9 and 21~meV in our calculation, which corresponds to a $\sim$0.45 meV/K peak shift. 
Our computed values are twice greater than the model and are in very good agreement with the $\sim$0.5 meV/K peak shift extracted from ARPES experiments on undoped cuprates~\cite{kim_anomalous_2002}.
\begin{figure}[t!]
\includegraphics[width=0.48\textwidth]{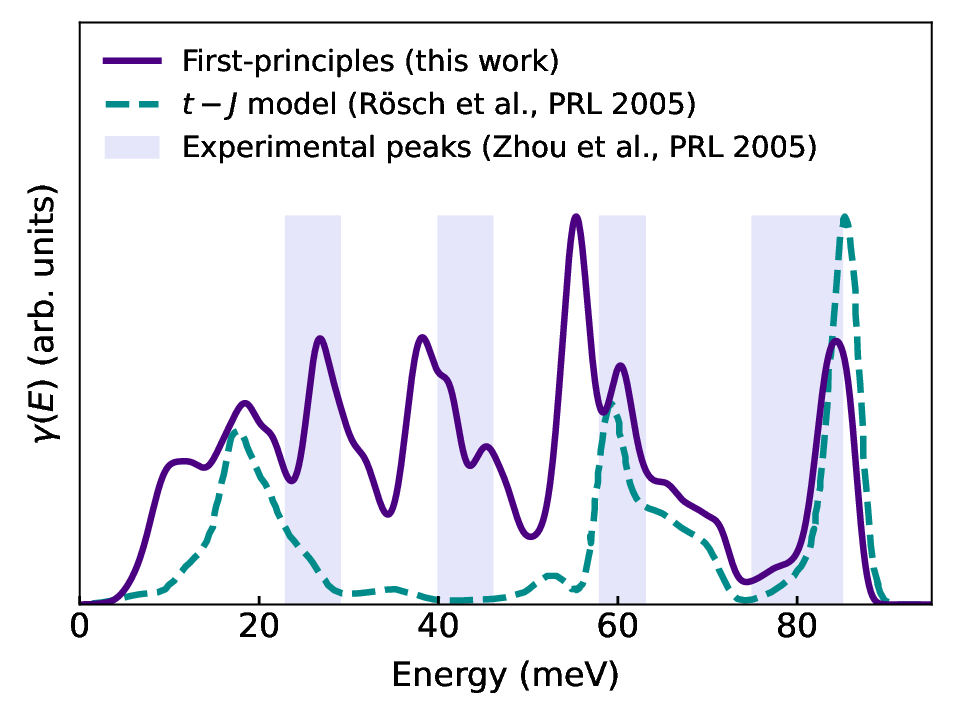}
\caption{Electron-phonon coupling distribution function, $\gamma(E)$, computed using Eq.~(\ref{eq:gamma}) (solid curve). For comparison, we show $t$–$J$ model results from Ref.~\cite{rosch_polaronic_2005} (dashed curve) and the range of the experimental self-energy peaks in LSCO  from Ref.~\cite{zhou_multiple_2005} (shaded regions). The distribution curves are normalized to the same maximum height.}
\label{fig:dist} 
\end{figure}
\\
\indent
Conventional analysis of $e$-ph interactions in cuprates focuses on the $e$-ph coupling distribution function~\cite{rosch_polaronic_2005}, which combines contributions to $e$-ph coupling from all phonons (with energies $\hbar \omega_{ \nu \mathbf{q} }$) at a given energy $E$: 
\begin{equation}
\label{eq:gamma}
\gamma(E)=\sum_{\nu\mathbf{q}}\left|g_{\nu}\left(\mathbf{k},\mathbf{q}\right)\right|^2\delta(E-\hbar \omega_{\nu\mathbf{q}}) \vspace{-5pt}.
\end{equation} 
This quantity has been computed using a $t$–$J$ model~\cite{rosch_polaronic_2005} to interpret the measured electron self-energy in underdoped LSCO~\cite{zhou_multiple_2005}. 
% explain the behavior
This model calculation gives a distribution function that captures the experimental self-energy peaks at 25, 60, and 80~meV (see Fig.~\ref{fig:dist}). However, the peak observed in experiments at $\sim$40~meV is absent in the model calculations. Previous work attributed this missing peak to surface effects or distortions due to doping in real samples~\cite{rosch_polaronic_2005}. 
\\
\indent
We compute the $e$-ph coupling distribution function for the nodal point $\mathbf{k}\!=\!N$ in our first-principles settings. 
As shown in Fig.~\ref{fig:dist}, our computed distribution function agrees with $t$–$J$ model results above 60~meV and brings the position of the 25-meV peak closer to experiments~\cite{zhou_multiple_2005}. Importantly, our calculations recover the missing peak at 40 meV, which is due to $e$-ph interactions with LE modes associated with bond-bending and apical O vibrations. 
% other than bond stretching
This demonstrates that the 40-meV feature observed in LSCO~\cite{zhou_multiple_2005} is already present in the undoped parent phase and is not a consequence of doping. 
The 40-meV feature is absent in the $t$–$J$ model because it considers only the Cu $d_{x^2-y^2}$ and O $p_{x,y}$ orbitals~\cite{rosch_electron-phonon_2004, zhang_effective_1988}, so $e$-ph interactions from Cu-O bond-bending and apical O vibrations are not properly described. 
Our results suggest that an accurate effective model of $e$-ph physics in cuprates would need to take into account additional electronic orbitals and phonon modes.
\\
\indent
In summary, we study $e$-ph interactions in the parent cuprate LCO using state-of-the-art first-principles calculations. We show that strong $e$-ph interactions are an intrinsic feature of the undoped phase and are mediated by two classes of LO phonons with Fr\"ohlich coupling, both consisting of Cu-O bond-bending, bond-stretching and apical O-atom vibrations, with smaller contributions from LE polar modes. 
Capturing this physics allows us to explain key features of the valence band spectral functions, including their significant broadening and QP weight renormalization, the presence of phonon sidebands with a broad incoherent background, and the origin of a 40~meV peak in the energy distribution functions not accounted for by existing models. 
As many parent high-temperature superconductors are ionic insulators, we believe that strong Fr\"ohlich-type \mbox{$e$-ph} coupling may be a general feature of parent phases, and plan to investigate this point in broader classes of superconductors in the future.

This work was primarily supported by the AFOSR and Clarkson Aerospace Corp under award FA9550-21-1-0460. Code development was supported by the National Science Foundation under Grant No. OAC-2209262. I.T. and N.M. acknowledge support by the NCCR MARVEL, a National Centre of Competence in Research, funded by the Swiss National Science Foundation (Grant number 205602). This research used resources of the National Energy Research Scientific Computing Center (NERSC), a U.S. Department of Energy Office of Science User Facility located at Lawrence Berkeley National Laboratory, operated under Contract No. DE-AC02-05CH11231.

\end{document}